\documentclass[prb,twocolumn,showpacs,amsmath,amssymb]{revtex4}

\usepackage{graphicx}
\usepackage{dcolumn}
\usepackage{bm}

\newcommand{\PFlowlevel}[1]{(TMTSF)$_2$#1}
\newcommand{\PF}{\PFlowlevel{PF$_6$}}           
\newcommand{\X}{\PFlowlevel{X}}                 
\newcommand{\myrefeq}[1]{Eq.\ (\ref{#1})}
\newcommand{\myreffig}[1]{Fig.~\ref{#1}}
\newcommand{\cs}[1]{\textsf{#1}}

\newcommand{\Omcminv}{$(\Omega\,\text{cm})^{-1}$}
\newcommand{\Tm}{T$^{-1}$}

\newcommand{\parl}[2]{\mv{#1}\|\mv{#2}}
\newcommand{\Tst}{T^\star}                          
\newcommand{\Tsdw}{T_{\text{SDW}}}
\newcommand{\mdeg}{^\circ}
\newcommand{\MR}{\Delta\rho/\rho_0}
\newcommand{\sncn}[1]{\sin^2\theta+\gamma_{#1}\cos^2\theta}
\newcommand{\ssncn}[1]{\sqrt{\sncn{#1}}}
\newcommand{\veps}{\varepsilon_0}
\newcommand{\De}{\Delta_1^2+\veps^2}
\newcommand{\pDe}{(\De)}
\newcommand{\tD}{\tilde{\Delta}}
\newcommand{\tv}{\tilde{v}}

\newcommand{\mv}[1]{\mathbf{#1}}                
\newcommand{\vek}[1]{$\mv{#1}$}                 
\newcommand{\mvk}{\mv{k}}
\newcommand{\vk}{\vek{k}}

\newcommand{\mvB}{\mv{B}}   \newcommand{\vB}{$\mvB$}
\newcommand{\mva}{\mv{a}}   \newcommand{\va}{$\mva$}
\newcommand{\mvb}{\mv{b'}}  \newcommand{\vb}{$\mvb$}
\newcommand{\mvc}{\mv{c^*}} \newcommand{\vc}{$\mvc$}
\newcommand{\mvj}{\mv{j}}   \newcommand{\vj}{$\mvj$}

\newcommand{\mja}{\parl{j}{a}}          \newcommand{\ja}{$\mja$}
\newcommand{\mjb}{\parl{j}{b'}}         \newcommand{\jb}{$\mjb$}
\newcommand{\mjc}{\parl{j}{c^\star}}    \newcommand{\jc}{$\mjc$}

\newcommand{\mBa}{\parl{B}{a}}          \newcommand{\Ba}{$\mBa$}
\newcommand{\mBb}{\parl{B}{b'}}         \newcommand{\Bb}{$\mBb$}
\newcommand{\mBc}{\parl{B}{c^\star}}    \newcommand{\Bc}{$\mBc$}

\newcommand{\plane}[2]{#1\text{--}#2}
\newcommand{\macplane}{\plane{\mva}{\mvc}}      \newcommand{\acplane}{$\macplane$}
\newcommand{\mbcplane}{\plane{\mvb}{\mvc}}      \newcommand{\bcplane}{$\mbcplane$}
\newcommand{\mabplane}{\plane{\mva}{\mvb}}      \newcommand{\abplane}{$\mabplane$}

\newcommand{\mBab}{\parl{\mvB}{(\mabplane)}}    \newcommand{\Bab}{$\mBab$}
\newcommand{\mBbc}{\parl{\mvB}{(\mbcplane)}}    \newcommand{\Bbc}{$\mBbc$}
\newcommand{\mBac}{\parl{\mvB}{(\macplane)}}    \newcommand{\Bac}{$\mBac$}

\newcommand{\Tfourlowlevel}[1]{$\Tst#1 4$~K}
\newcommand{\Tfour}{\Tfourlowlevel{\approx}}        

\begin{document}

\preprint{Version: \today}

\title{Unconventional Spin Density Wave in \PF\ below \Tfour}

\author{Mario Basleti\'{c}}
 \email{basletic@phy.hr}
\affiliation{Department of Physics, Faculty of Science, POB 331, HR-10002 Zagreb, Croatia}

\author{Bojana Korin-Hamzi\'{c}}
\affiliation{Institute of Physics, POB 304, HR-10001 Zagreb, Croatia}

\author{Kazumi Maki}
\affiliation{Department of Physics, University of Southern California, Los Angeles CA 90089-0484, USA\\
            and\\
            Max-Planck Institute for the Physics of Complex Systems, N\"{o}thnitzer Str.38, D-01187 Dresden,
            Germany}
\date{\today}

\begin{abstract}
The presence of subphases in spin-density wave (SDW) phase of \PF\ below \Tfour\ has been suggested by
several experiments but the nature of the new phase is still controversial. We have investigated the
temperature dependence of the angular dependence of the magnetoresistance in the SDW phase which shows
different features for temperatures above and below \Tfour. For $T > 4$~K the magnetoresistance can be
understood in terms of the Landau quantization of the quasiparticle spectrum in a magnetic field, where
the imperfect nesting plays the crucial role. We propose that below \Tfour\ the new unconventional SDW
(USDW) appears modifying dramatically the quasiparticle spectrum. Unlike conventional SDW the order
parameter of USDW depends on the quasiparticle momentum $\Delta_1(\mvk)\propto \cos 2bk_y$. The present
model describes many features of the angular dependence of magnetoresistance reasonably well. Therefore,
we may conclude that the subphase in \PF\ below \Tfour\ is described as SDW plus USDW.
\end{abstract}


\maketitle

\section{Introduction}
The very anisotropic organic conductors \X\ (where TMTSF is tetramethyltetraselenafulvalene and X =
PF$_6$, AsF$_6$, ClO$_4$ \dots\ stands for monovalent anion) or Bechgaard salts continue to attract much
attention since the discovery of their superconductivity in 1979.\cite{JeromeJP80} A variety of
electronic ground states under pressure and/or magnetic field, (conventional) spin density wave (SDW),
field induced spin density wave (FISDW) with quantum Hall effect and unconventional (most likely
$p-$wave) superconductivity, are very intriguing.\cite{IshiguroBook98, LeePRL02}

\PF\ is metallic down to $\Tsdw\approx 12$~K, where the transition into the semiconducting SDW state
occurs. It is known that SDW in \PF\ undergoes another transition at $\Tst\approx\Tsdw/3$ (at 3.5--4~K
at ambient pressure).\cite{TakahashiJPSJ86, TakahashiSM91, LasjauniasPRL94} The indication of the
subphase transition was first seen by NMR,\cite{TakahashiJPSJ86} where $T^{-1}_1$ diverges and the spin
susceptibility changes at $\Tst$. The transition at $\Tst$ is preserved through the entire $P-T$ phase
diagram. Furthermore, a calorimetric transition at 3.5~K, with a large hysteretic phenomena in the
temperature range 2.5--4~K (caused by the sample history), has been observed and interpreted as an
indication of a glass transition.\cite{LasjauniasPRL94} On the other hand, the low frequency dielectric
relaxation of SDW in \PF\ did not show the existence of the glass transition.\cite{TomicSM97} Since
then, the SDW state was widely investigated, but the nature of the possible subphases remains
controversial. Our study of the angular dependence of the magnetoresistance (MR) for \Bab\ plane has
shown dramatically different features above and below \Tfour.\cite{BasleticSM99, KorinHamzicJP99}
However, taking into account our MR results for temperatures $T \geq 2.2$~K, the transition at $\Tst$
appears to be unique to \PF, as it has not been identified for X = AsF$_6$ and
ClO$_4$.\cite{KorinHamzicSM01} On the other hand, there are a few
reports\cite{KagoshimaSSC99,LasjauniasEJB99} indicating similar transition in \PFlowlevel{AsF$_6$},
though less pronounced than in \PF. Therefore, at this moment, we cannot exclude the presence of similar
transitions in other Bechgaard salts.

Recently, we have studied the MR in \PF, with a magnetic field rotated within the \acplane\ plane, which
behaves differently for $T>4$~K and $T<4$~K at ambient pressure.\cite{KorinHamzicEPL98} For $T>4$~K the
magnetoresistance was described in terms of the quasiparticles scattered by the \vk\ dependent
scattering rate (where \vk\ is the quasiparticle wave vector). In other words, we could understand the
magnetotransport in terms of the standard Fermi liquid theory, i.e.\ by the quasiparticles with the
energy gap given in the model with imperfect nesting.\cite{HuangPRB89} In spite of the fact that for
$T<4$~K we had to introduce a rather artificial scattering rate $\Gamma(\phi=bk_y)$ the description of
the resistance along the \vb\ axis was not satisfactory.\cite{KorinHamzicEPL98}

More recently, an unconventional density wave (USDW and UCWD) was proposed as a possible ground state of the
electronic systems in organic conductors and heavy fermion
systems.\cite{DoraEJB167,NersesyanJPCM91,IkedaPRL98,BenfattoEPJB,ChakravertyPRB01} Unlike the conventional
SDW, the USDW is defined as the SDW where the order parameter $\Delta(\mvk)$ depends on the quasi-particle
momentum \vk. In spite of a clear thermodynamic signal (as in the usual mean field-like transition), the
first order term in $\Delta(\mvk)$, corresponding to local charge or local spin, is invisible. Consequently,
these states may be called the phase with hidden order parameter.\cite{ChakravertyPRB01} UCWD has been
identified very recently, from the temperature dependence of the threshold electric field,\cite{BasleticSM01}
in the low temperature phase of $\alpha$-(ET)$_2$KHg(SCN)$_4$.\cite{DoraPRB01} Similarly, a mysterious
micromagnetism seen in AF phase of URu$_2$Si$_2$ could also be interpreted in terms of
USDW.\cite{VirosztekCONDMAT0112}

The aim of this work was to see if the presence of possible subphases in the SDW below 4~K could be
observed in the temperature dependence of the conductivity and MR as well as in the anisotropy of the
MR. In this paper we compare the experimental MR data of \PF\ in the SDW state, showing the pronounced
differences for $T>4$~K and $T<4$~K, with our new theoretical results (preliminary results in Ref.\
\onlinecite{KorinHamzicIJMPB01}). We propose that the anomaly at \Tfour\ in \PF\ signals the appearance
of USDW. We point out that USDW requires more subtle balance between different interaction terms than
conventional SDW,\cite{DoraEJB167} and consequently it is perhaps not easily find in other Bechgaard
salts.

\section{Experiment.}
The measurements were done down to 2~K in magnetic fields up to 5~T and with different directions of the
current (through the crystal) and different orientations of magnetic field. A rotating sample holder
enabled the sample rotation around a chosen axis over a range of $190\mdeg$. The single crystals used
come all from the same batch. Their \va\ direction is the highest conductivity direction, the \vb\
direction (with intermediate conductivity) is perpendicular to \va\ in the $\plane{\mva}{\mv{b}}$ plane
and \vc\ direction (with the lowest conductivity) is perpendicular to the $\plane{\mva}{\mv{b}}$ (and
\abplane). The room temperature conductivity values for $\sigma_\mva$, $\sigma_\mv{b}$, and
$\sigma_\mv{c}$ are 500~\Omcminv, 20~\Omcminv, and 1/35~\Omcminv\ respectively.

The experimental MR data, that will be analyzed here, are for \vc\ and \vb\ axis and for different
orientations of magnetic field. The MR, defined as $\MR = [\rho(B)-\rho(0)]/\rho(0)$, was measured in
various four probe arrangements on samples cut from a long crystal. Moreover, the measurements of \vc\
axis MR, for two different magnetic field rotations, were performed on the same sample but which was cut
to two parts. In the case of $\rho_\mv{b}$ (\jb) two pairs of the contacts were placed on the opposite
\acplane\ surfaces while for $\rho_\mv{c}$ (\jc) on the opposite \abplane\ surfaces. We used very slow
cooling rates (about 2--5~K per hour) to avoid the appearance of the irreversible resistance jumps
usually encountered for \ja\ measurements. This was especially important for \jb\ geometry, where
additional care was required to avoid possible mixture of $\sigma_\mv{b}$ and $\sigma_\mv{c}$
conductivities.\cite{KorinHamzicEPL98} This can be described by using the concept of the equivalent
isotropic sample that gives a simple picture of the current distribution in the anisotropic
sample.\cite{CooperMCLC85} The eligible test for properly measured \vb-axis resistivity is linear
temperature dependence at high temperatures.\cite{BechgaardSSC80} Namely, there is a non-monotonic
temperature dependence of $\rho_\mv{c}$ in \PF\ at ambient pressure going through a well characterized
maximum at 80~K in contrast to the results for $\rho_\mv{a}$ ($\propto T^2$) and $\rho_\mv{b}$ ($\propto
T$) exhibiting a monotonous, metallic-like decrease upon lowering temperature.

\begin{figure}
\includegraphics*[width=8.5cm]{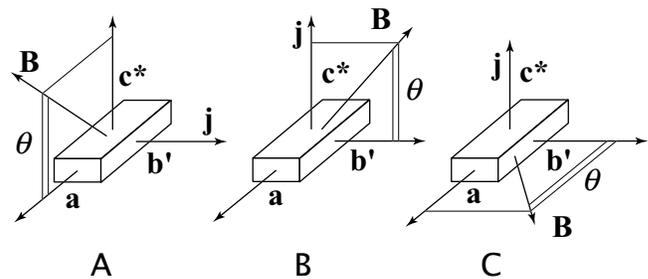}
\caption{\label{fig:config} Three configurations (case \cs{A}, \cs{B}\, and \cs{C}) of the current \vj\
and magnetic field \vB\ direction. (See text for a detailed explanation.)}
\end{figure}

Figure \ref{fig:config}\ present three configurations that will be analyzed in this work: a)
\myreffig{fig:config}\cs{A} shows the case when the current direction is along the \vb\ axis and the
magnetic field is rotated in the \acplane\ (\jb, \Bac) perpendicular to the current direction. $\theta$
is the angle between \vB\ and the \va\ axis, i.e.\ $\theta=0$ for \Ba\ and $\theta=90\mdeg$ for \Bc. b)
\myreffig{fig:config}\cs{B} shows the case when the current direction is along the \vc\ axis and the
magnetic field is rotated in the \bcplane\ plane (\jc, \Bbc). $\theta$ is the angle between \vB\ and the
\vb\ axis, i.e. $\theta=0$ for \Bb\ and $\theta=90\mdeg$ for \Bc. c) \myreffig{fig:config}\cs{C} shows
the case when the current direction is along the \vc\ axis and the magnetic field is rotated in the
\abplane\ plane (\jc, \Bab) perpendicular to the current direction. $\theta$ is the angle between \vB\
and the \vb\ axis, i.e. $\theta=0$ for \Bb\ and $\theta=90\mdeg$ for \Ba.

\section{Model, results and discussion.}
\subsection{Quasiparticle spectrum above \Tfour.}
We limit ourselves to the \vc\ axis magnetoresistance, i.e.\ to the case \cs{B} when the current \vj\
direction is along the \vc\ axis, the magnetic field is rotated in the \bcplane\ plane with $\theta =
\measuredangle(\mvb, \mvB)$. We leave a detailed analysis for another current directions and magnetic
field orientations above 4~K for a future publication.

The Landau quantization of the quasiparticle spectrum appears to describe very well the observed
results. In the limit of perfect nesting all the electron orbits are open and there will be no Landau
quantization. On the other hand, in the presence of the imperfect nesting\cite{HuangPRB89} as in \PF\,
the quasiparticle energy landscape develops local minima at $k_z=\pm k_F$, $k_y=\pm\pi/2b$. In other
words, closed orbits appear and they will be quantized in the presence of a magnetic field.

For $T>4$~K the quasiparticle energy is given by:\cite{KorinHamzicEPL98}
\begin{eqnarray}
E_\mvk & = &       \sqrt{\eta^2+\Delta^2}-\veps\cos 2bk_y \nonumber \\
        & \approx & \Delta - \veps + \frac{1}{2\Delta}\eta^2 + 2\veps b^2k_y^2\,, \label{eq:QPEnergyAbove}
\end{eqnarray}
where $\eta=\left[ v_a^2(k_x-k_F)^2 + v_c^2k_z^2 \right]^{1/2}$ is the quasiparticle energy in the
normal state ($v_a$ and $v_c$ are Fermi velocities in \va\ and \vc\ direction, respectively), $\Delta$
($\approx 34$~K) is the order parameter for conventional SDW and $\veps$ ($\approx 13$~K) is the
parameter characterizing the imperfect nesting effect.\cite{KorinHamzicEPL98} The quasiparticle energy
is expanded for small $(k_x-k_F)^2$ and $k_y^2$. In a presence of a magnetic field within the \bcplane\
plane, with $\theta$ being the angle between the magnetic field \vB\ and the \vb\ axis, the minimum
energy (i.e. the energy gap) in \myrefeq{eq:QPEnergyAbove}\ is given by:
\begin{equation}\label{eq:EnergyGapAbove}
E(B,\theta) \approx \Delta - \veps + \sqrt{\frac{\veps}{\Delta}}\,v_a b e B\ssncn{2}\,,
\end{equation}
with $\gamma_2=(1/\veps\Delta)(v_c/2b)^2$. For $B=0$ the resistance is given as $\rho_{zz}(T,0)\propto
\exp[\beta E(0,\theta)]$, whereas for $B\neq0$ we have:
\begin{equation} \label{eq:MRzzAboveProp}
\rho_{zz}(T,B) \propto B\ssncn{2}\,e^{\beta E(B,\theta)}\,.
\end{equation}
First, we note that the energy gap is given in the both limits ($B=0$ and $B\neq 0$) by
\myrefeq{eq:EnergyGapAbove}. Second, for $\omega_c\tau>1$, where $\omega_c$ is the cyclotron frequency and
$\tau$ is the scattering rate, $\rho_{zz}(T,B)$ contains a $B$ linear coefficient. So, we may interpolate
these expressions as:
\begin{eqnarray} \label{eq:MRzzAbove}
\rho_{zz}(B,T) & \approx & 
                        e^{ \beta (\Delta-\veps) \left( 1+A_2B\ssncn{2}\,\right) } \nonumber \\
   & &               \times\left( 1+C_2B\ssncn{2} \right)\,,
\end{eqnarray}
with $A_2 = (\veps/\Delta)^{1/2} v_a b e/(\Delta-\veps)$.

\begin{figure}
\includegraphics*[width=7cm]{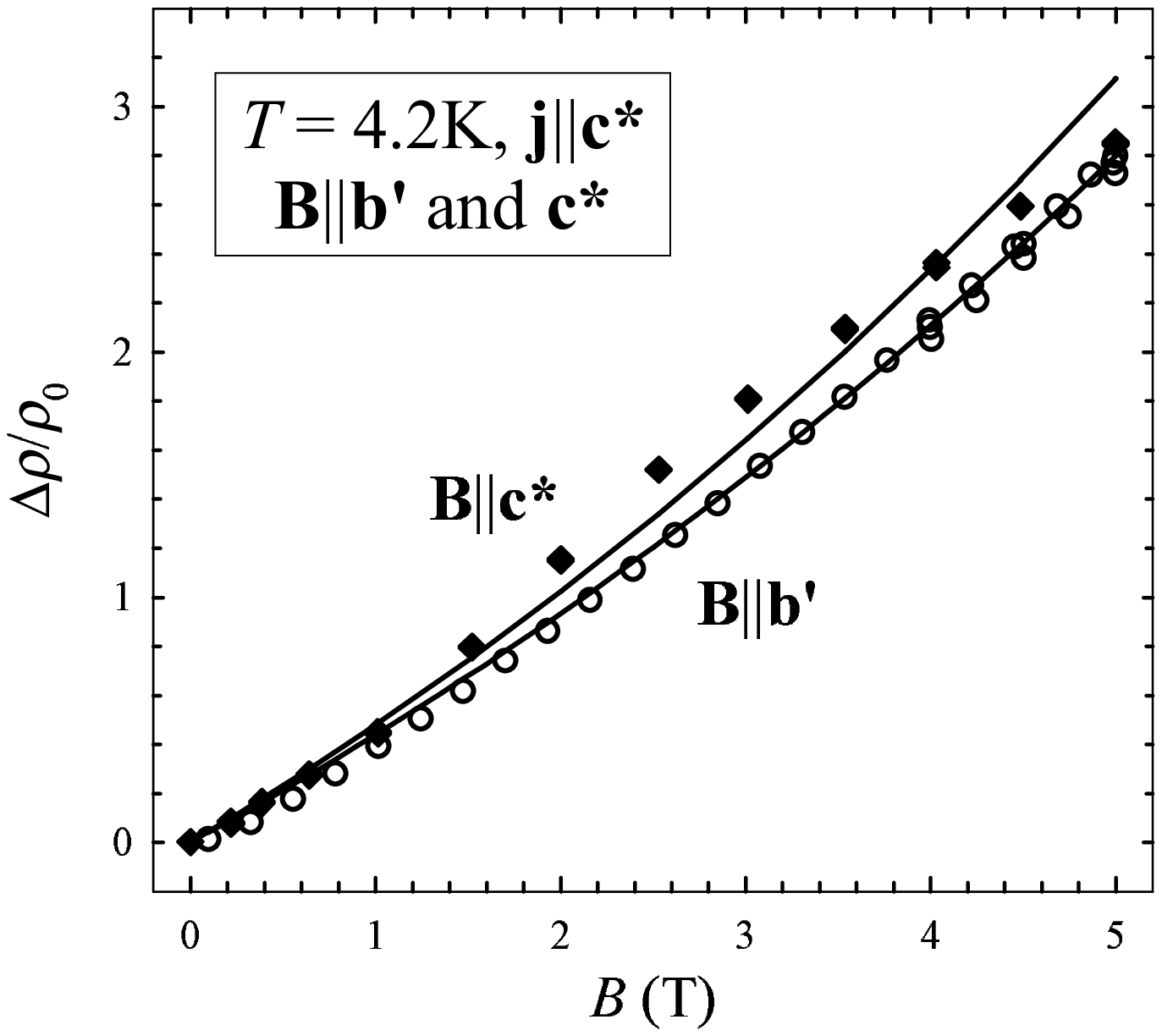}
\caption{\label{fig:MRvsBAbove} Magnetic field dependence of $\MR$ at 4.2~K for \jc, \Bb\ and \Bc. Solid
lines are fits to the theory (see text).}
\end{figure}
\begin{figure}
\includegraphics*[width=7cm]{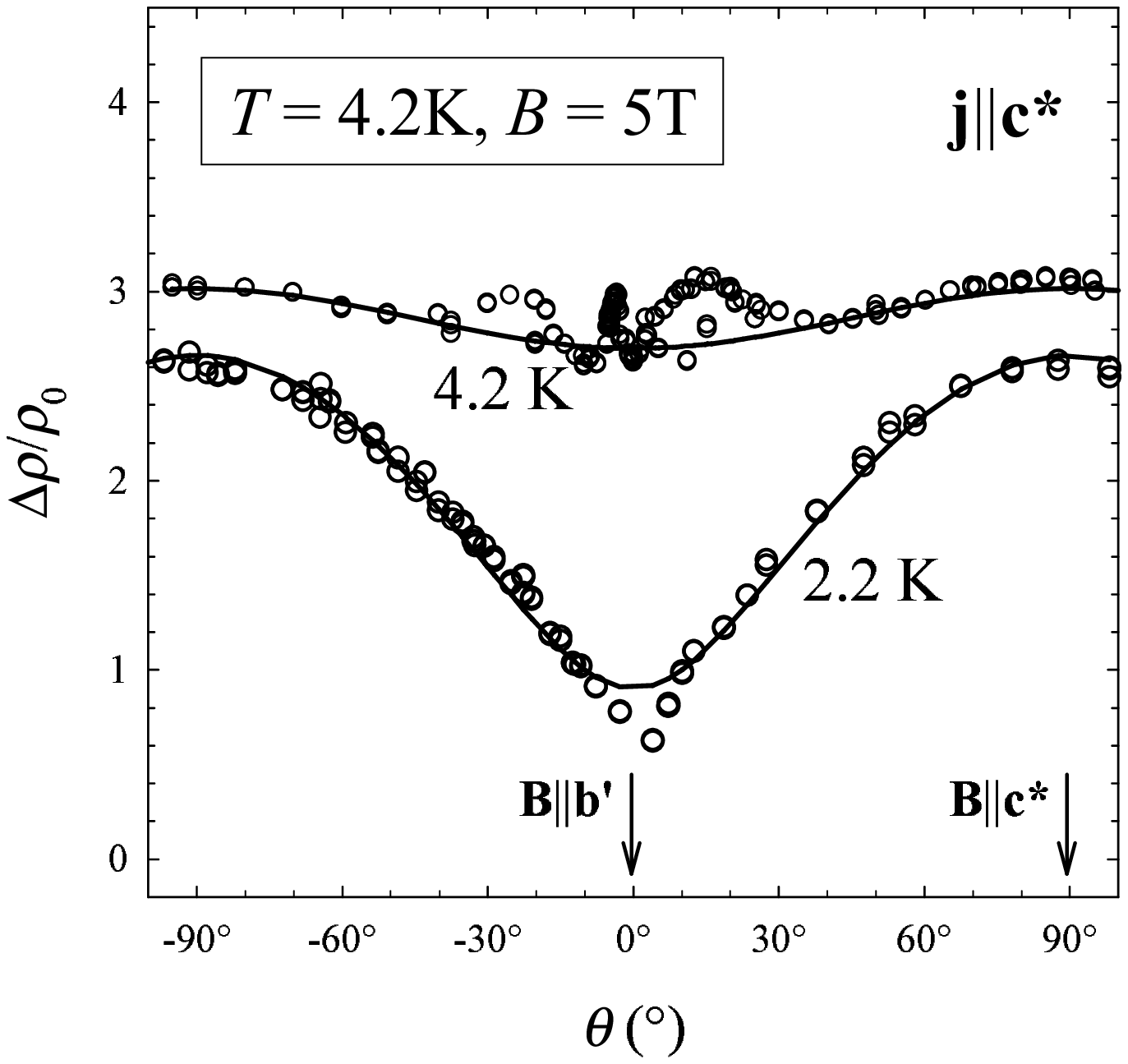}
\caption{\label{fig:MRvsThAbove} Angular dependence of $\MR$ at 2.2~K and 4.2~K, $B=5$~T, for \jc, \vB\ in
\bcplane\ plane. Solid lines are fits to the theory (see text).}
\end{figure}
We shall compare now our experimental data with the above equations. The magnetic field dependence of MR
for \jc, \Bb\ and \Bc\ at 4.2~K is presented in \myreffig{fig:MRvsBAbove}. \myreffig{fig:MRvsThAbove}
shows the angular dependence of MR for \jc, $B=5$~T at 4.2~K and 2.2~K. $\theta$ is the angle between
\vB\ and the \vb\ axis (see \myreffig{fig:config}, case \cs{B}). Solid line is the fit based on the
\myrefeq{eq:MRzzAbove}. The change in the angular dependence of MR for $T>4$~K and $T<4$~K is clearly
seen (the case for $T=2.2$~K will be treated in Section \ref{sec:BelowTstar}). It is evident that the
present model describes rather well the data, with both the field and angular dependence of MR, at
$T=4.2$~K and $B=5$~T giving $(\Delta-\veps)=21$~K, $A_2=0.014$~\Tm, $C_2=0.38$~\Tm\, and
$\gamma_2=0.85$. These values enable us to extract the \va\ axis coherence length $\xi_a = v_a/\Delta
=1.2 \times 10^{-6}$~cm and $v_c/v_a = 7.33 \times 10^{-2}$ (we used here flux quantum $\pi/e = \Phi_0 =
2.07 \times 10^{-11}$~T$\,$cm$^2$). Both $\xi_a$ and $v_c/v_a$ thus deduced are consistent with the ones
deduced from the anisotropy in the resistivity.\cite{IshiguroBook98}

\subsection{\label{sec:BelowTstar}Quasiparticle spectrum below \Tfour.}
We start by proposing that the anomaly at \Tfour\ in
\PF\ signals the appearance of USDW with the momentum dependent order parameter $\Delta_1(\mvk)=\Delta_1
\cos 2\phi$, where $\phi=bk_2$ with wave vector $\mv{Q}=(2k_F,\pi/2b,0)$. In other words, below $\Tst$
two order parameters (SDW and USDW) coexist. In this case the quasiparticle spectrum changes from
\myrefeq{eq:QPEnergyAbove}\ to:
\begin{widetext}
\begin{eqnarray}
E_\mvk & = &    \sqrt{
                    \left(
                            \sqrt{\eta^2+\Delta^2}-\veps\cos 2bk_y
                    \right)^2 + \Delta_1^2 \cos^2 2\phi
                } \label{eq:QPEnergyBelowExact} \\
       & = &    \sqrt{
                    (\eta^2+\Delta^2) \frac{\Delta_1^2}{\De}
                    +\pDe \left( \cos^2 2\phi - \cos^2 2\phi_0 \right)^2
                } \nonumber \\
 & \approx &    \sqrt{
                    \tD^2 + \tv_a^2(k_x-k_F)^2 + \tv_c^2k_z^2
                    + 4 \pDe b^2k_y^2
                }\,, \label{eq:QPEnergyBelow}
\end{eqnarray}
\end{widetext}
where $\tD^2=\Delta \Delta_1 \pDe^{-1/2}$, $\tv_a = v_a\Delta_1 \pDe^{-1/2}$ and $\tv_c = v_c \Delta_1
\pDe^{-1/2}$. We have not included a constant shift in $k_y$ and $k_z$, since they are of no importance
when one considers the effect of the magnetic field. In the absence of the magnetic field, the effect of
$\Delta_1$ (or USDW) is to change the minimum energy gap from $E_{\text{min}} = \Delta-\veps$ ($T>4$~K)
to $E_{\text{min}} =\tD$ ($T<4$~K). As we shall see later, the introduction of the magnetic field
changes dramatically the minimum energy gap $E_{\text{min}}$. Such a dramatic shift in $E_{\text{min}}$
in USDW and UCDW in a magnetic field has already been discussed in Ref.\ \onlinecite{NersesyanJLTP89}
and \onlinecite{NersesyanJPCM91}.

We shall see in the following that the field and the angle dependent quasiparticle spectrum describes
the angle dependent MR observed in \PF\ below \Tfour\ rather satisfactory. The quasiparticle energy gap
in the absence of magnetic field is given by Eqs.\ (\ref{eq:QPEnergyBelowExact}) and
(\ref{eq:QPEnergyBelow}). Due to the quadratic form in \vk\ in the square root, we expect the Landau
quantization in the presence of magnetic field. Let us consider three cases (\myreffig{fig:config})
separately.

\subsubsection{Case \cs{A}: \jb, \Bac, $\theta = \measuredangle (\mva,\mvB)$}
We can recast \myrefeq{eq:QPEnergyBelow}\ as an eigenvalue problem:
\begin{eqnarray}
E^2(B,\theta) \psi & = & \bigg[
                    \tD^2 + \tv_a^2(eBy\cos\theta)^2 + \tv_c^2(eBy\sin\theta)^2 \nonumber \\
& &                    - (2b)^2\pDe\frac{\text{d}^2}{\text{d}y^2} \bigg] \psi\,,
\end{eqnarray}
where $\psi$ is the electron wave function. This gives readily for the quasiparticle energy corresponding to
the $n-$th Landau level:
\begin{equation}
E^2_n(B,\theta) = \tD^2 + 2\tv_a\Delta_1 e B (\sncn{1})^{1/2}(2n+1)\,,
\end{equation}
($n=0,1,2\ldots$). From this we obtain the minimum energy gap $E_{\text{min}}$:
\begin{equation}
    E_{\text{min}}(B,\theta) = \tD \sqrt{1+A_1|B|(\sncn{1})^{1/2}}\,,
\end{equation}
\begin{equation}
A_1 = \frac{2\tv_a\Delta_1}{\tD^2}be\,,\ \gamma_1 =
  \left( \frac{\tv_c}{\tv_a} \right)^2 \sim 10^{-3}\,. \nonumber
\end{equation}
In this configuration $\gamma_1$ is clearly negligible. By approximating the cyclotron frequency as:
\begin{eqnarray}
E_1(B,\theta)-E_0(B,\theta) = \nonumber \\
= \tD \left(
    \sqrt{1+3A_1|B\sin\theta|} - \sqrt{1+A_1|B\sin\theta|}
    \right) \nonumber \\
\approx \tD A_1 |B\sin\theta|\,,
\end{eqnarray}
and noting the fact that in the presence of magnetic field
\[
\sigma_{yy} \approx
    \frac{ |B\sin\theta| \exp
            \left[
                -\beta\Delta\sqrt{ 1 + A_1 |B\sin\theta| }
            \right]
         }
         {
            1 + C' \left( B\sin\theta \right) ^2
         }\,,
\]
we finally obtain the interpolation formula:
\begin{eqnarray} \label{eq:MRyyBac}
\rho_{yy} & \approx & \exp \left( \beta\tD\sqrt{1+A_1|B\sin\theta|} \right) \nonumber \\
 & &\times\left(
        1 + C_1 A_1|B\sin\theta|
    \right)\,,
\end{eqnarray}
where $C_1 = (\tD/\Gamma)^2$ and $\Gamma$ is the quasiparticle relaxation rate (note that $\Gamma$ is
\vk-independent).

\begin{figure}
\includegraphics*[width=7cm]{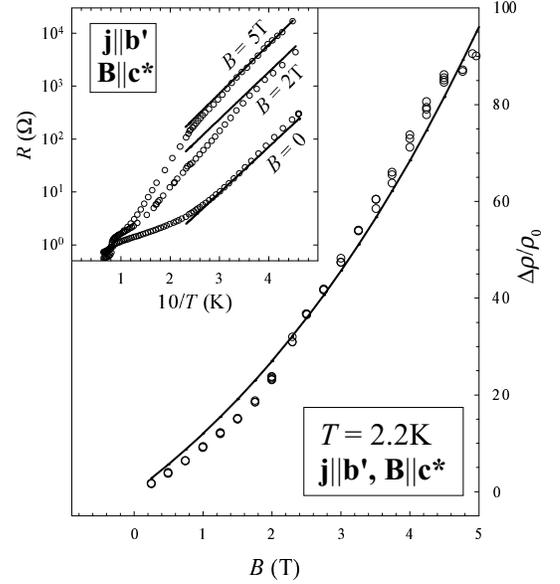}
\caption{\label{fig:MRyyBac} Magnetic field dependence of $\MR$ at 2.2~K for \jb\ and \Bc. Inset: $R$
vs.\ inverse temperature for $B=0$, 2~T, and 5~T. Solid lines are fits to the theory (see text).}
\end{figure}
\begin{figure}
\includegraphics*[width=7cm]{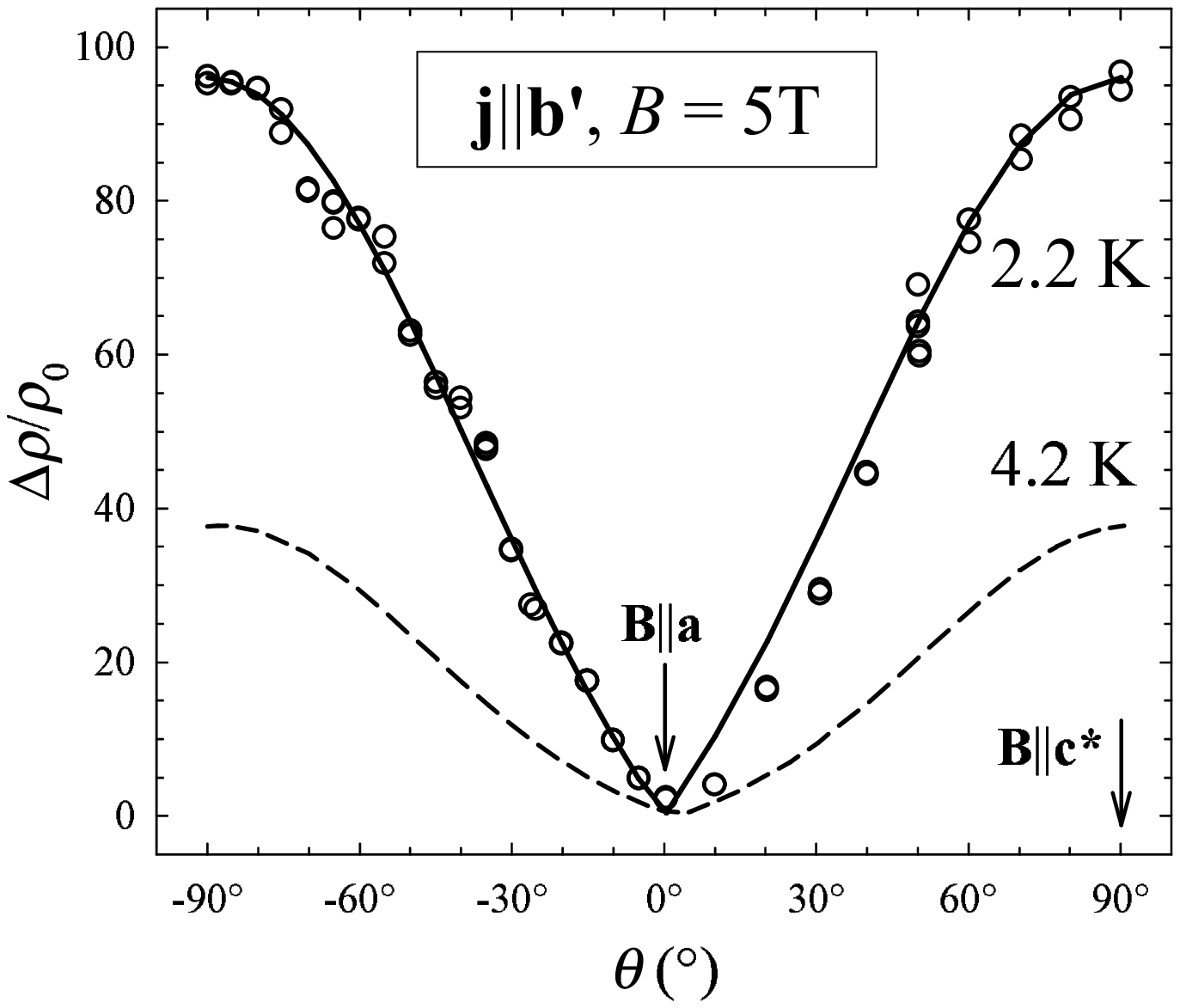}
\caption{\label{fig:MRyyThac} Angular dependence of $\MR$ at 2.2~K and 4.2~K, $B=5$~T, for \jb, \vB\ in
\acplane\ plane. Solid line is fit to the theory (see text).}
\end{figure}
The comparison of \myrefeq{eq:MRyyBac}\ (with $\theta=\pi/2$) with the experimental data is given in
Figs.\ \ref{fig:MRyyBac} and \ref{fig:MRyyThac}. \myreffig{fig:MRyyBac} shows the results of the
magnetic field dependence of the MR at 2.2~K for \jb\ and \Bc. The inset shows the temperature
dependence of the MR for $B=5$~T in the same geometry. The solid lines show the fit to the theoretical
model explained above. \myreffig{fig:MRyyThac} shows the angular dependence of MR for \jb\ and $B=5$~T
at 2.2~K (see \myreffig{fig:config}, case \cs{A}). Dashed line shows the results at 4.2~K. Solid line is
fit based on \myrefeq{eq:MRyyBac}. Further, the $1/T$ dependent magnetoresistance is compared in the
inset of \myreffig{fig:MRyyBac}. By fitting the data we can deduce $\tD=20$~K, $A_1 = 0.027$~\Tm\ which
gives $\Delta_1/\Delta=0.568$ (where we took $b = 0.77$~nm and $\xi_a=\tv_a/\tD = 120$~\AA). We obtain
the USDW order parameter $\Delta_1\approx 20$~K. These numbers look rather reasonable. So, in this
geometry, the present model describes the experimental data reported in Ref.\
\onlinecite{KorinHamzicEPL98} rather well.

\subsubsection{Case \cs{B}: \jc, \Bbc, $\theta = \measuredangle (\mvb,\mvB)$}
In this configuration the eigenequation is rewritten as:
\begin{eqnarray}
E^2(B,\theta) \psi & = & \bigg[
              \tD^2 - \tv_a^2 \frac{\text{d}^2}{\text{d}y^2}+ \tv_c^2(eBx\cos\theta)^2 \nonumber \\
& &                   + (2b)^2\pDe (eBx\sin\theta)^2  \bigg] \psi\,,
\end{eqnarray}
which is solved as:
\begin{equation}
E^2_n(B,\theta) = \tD^2 + 2\tv_a\Delta_1 e B (\sncn{2})^{1/2}(2n+1)\,,
\end{equation}
($n=0,1,2\ldots$). Therefore, the minimum energy gap $E_{\text{min}}$ is:
\begin{equation} \label{eq:EminBbc}
E_{\text{min}}(B,\theta) = \tD \sqrt{1+A_2|B|(\sncn{2})^{1/2}}\,,
\end{equation}
\begin{equation}
A_2 = \frac{2\tv_a\Delta_1}{\tD^2}be\,,\ \gamma_2 =
      \frac{\tv_c^2}{(2b)^2 \pDe}\,. \nonumber
\end{equation}
The magnetoresistance along the \vc\ axis is given by:
\begin{eqnarray} \label{eq:MRzzBbc}
\rho_{zz} & \approx & \exp\left[ \beta\tD\sqrt{1+A_2|B|(\sncn{2})^{1/2}} \right]  \nonumber \\
& &    \times \left( 1 + C_2 A_2|B|\ssncn{2} \right)\,.
\end{eqnarray}

\begin{figure}
\includegraphics*[width=7cm]{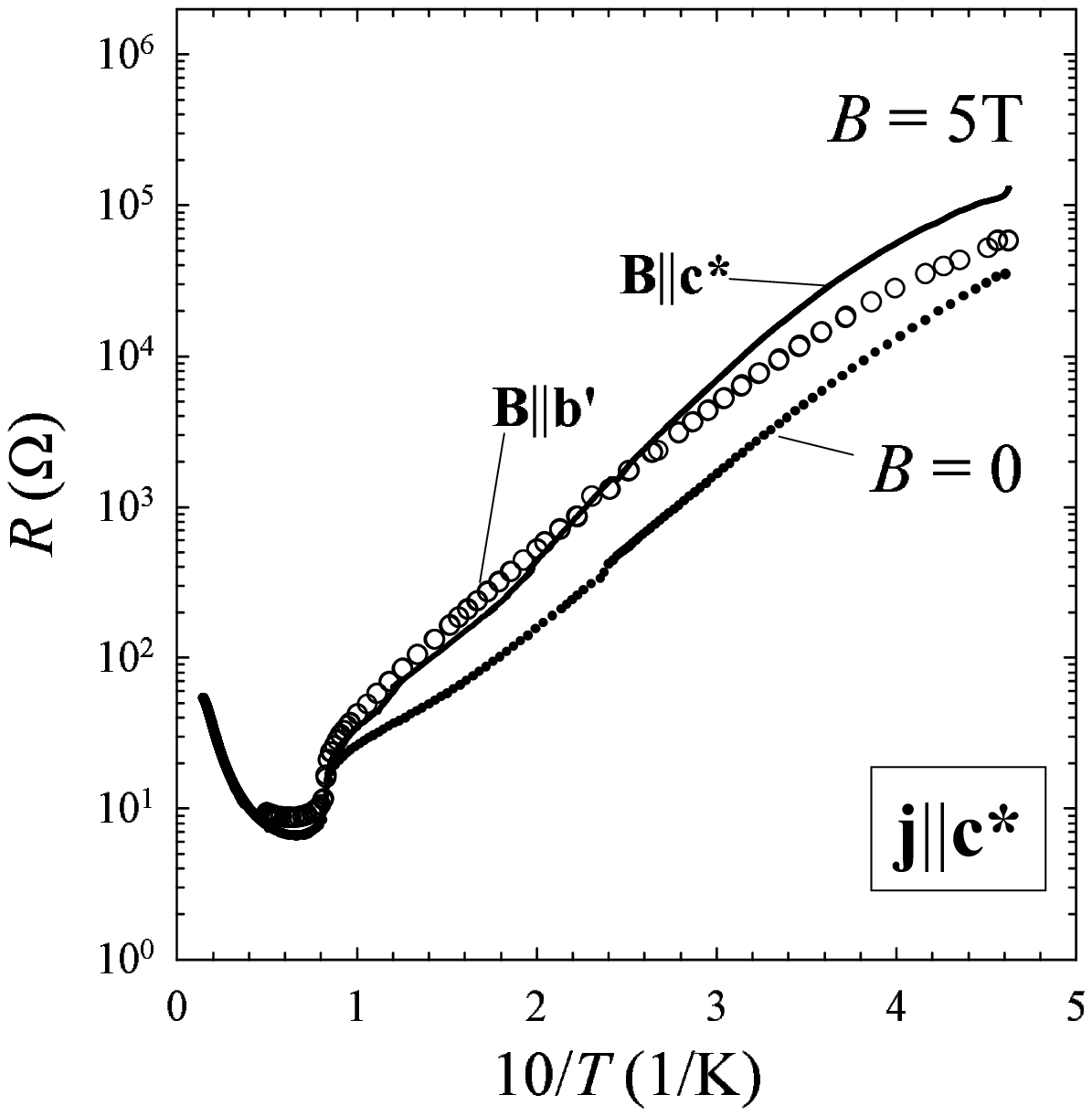}
\caption{\label{fig:RzzBbc} Temperature dependence of the resistance $R$ for \jc, $B=0$ and $B=5$~T (for two
different magnetic field orientations \Bb\ and \Bc).}
\end{figure}
\begin{figure}
\includegraphics*[width=7cm]{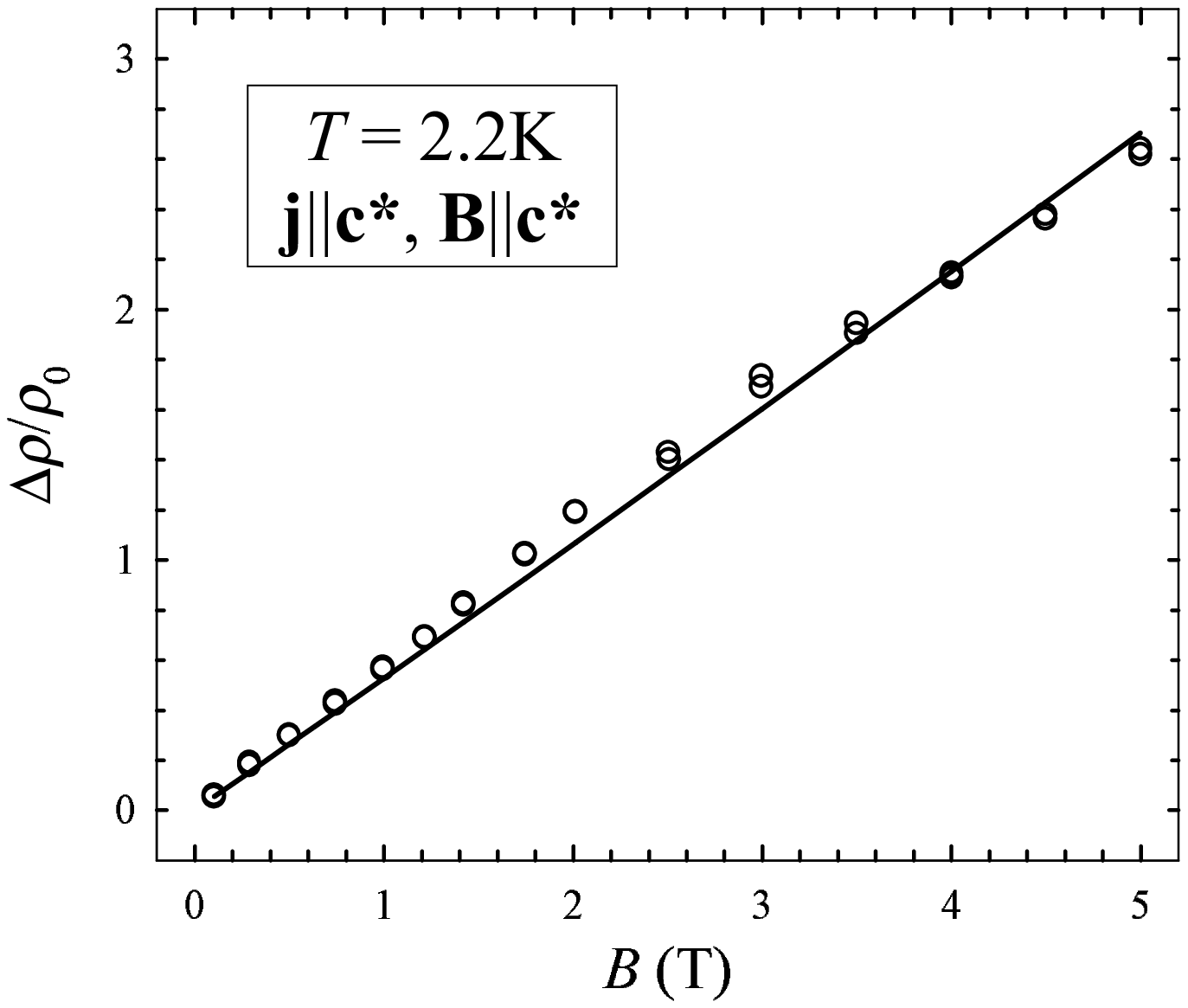}
\caption{\label{fig:MRzzBbc} Magnetic field dependence of $\MR$ at 2.2~K for \jc\ and \Bc. Solid line is fit
to the theory (see text).}
\end{figure}
Figure \ref{fig:RzzBbc} shows the temperature dependence of the resistance for \jc\ ($B=0$, $B=5$~T) and
for two different magnetic field orientations \Bb\ and \Bc. The difference in $R$ vs.\ $10/T$ behaviour
below $\approx 4$~K for two magnetic field orientations is clearly observed. The magnetic field
dependence of magnetoresistance for \jc\ and \Bc\ at $2.2$~K is presented in \myreffig{fig:MRzzBbc}. As
mentioned before, \myreffig{fig:MRvsThAbove}\ shows also the angular dependence of magnetoresistance for
\jc, $B=5$~T at 2.2~K (see \myreffig{fig:config}, case \cs{B}). Solid line (on both
\myreffig{fig:MRvsThAbove} and \myreffig{fig:MRzzBbc}) is fit based on the \myrefeq{eq:MRzzBbc}. The
present expression is comparable with both the $B$ dependence of magnetoresistance for $\theta=\pi/2$
and the $\theta$ dependent magnetoresistance at $T = 2.2$~K and $B = 5$~T. Again, the present model
describes the data rather well. In the present comparison we deduce $A_2 = 0.00134\text{~\Tm}=
A_1/20.2$, $C_2 = 0.5192\text{~\Tm} = C_1/20.2$, and $\gamma_2=0.060$ which gives $\tD =20$~K and
$v_c/v_a = 0.02$. On the other hand, we obtain $\Delta_1/\Delta=0.0284$ that gives $\Delta_1\approx
1$~K. This implies the USDW order parameter in the present configuration is reduced by a factor of 1/20
compared with the one in the first configuration. This result is rather unexpected, but we hope the
future work will clarify this problem.

\subsubsection{Case \cs{C}: \jc, \Bab, $\theta = \measuredangle (\mvb,\mvB)$}
In this configuration the eigenequation is rewritten as:
\begin{eqnarray}
E^2(B,\theta) \psi & = & \bigg[
              \tD^2 - \tv_c^2 \frac{\text{d}^2}{\text{d}z^2}+ \tv_a^2(eBz\sin\theta)^2 \nonumber \\
& &                   + (2b)^2\pDe (eBz\cos\theta)^2  \bigg] \psi\,,
\end{eqnarray}
which gives
\begin{equation}
E^2_n(B,\theta) = \tD^2 + 2\tv_c\tv_a e B (\sncn{3})^{1/2}(2n+1)\,,
\end{equation}
($n=0,1,2\ldots$). For the minimum energy gap $E_{\text{min}}$ we obtain:
\begin{equation} \label{eq:EminBab}
E_{\text{min}}(B,\theta) = \tD \sqrt{1+A_3|B|(\sncn{3})^{1/2}}\,,
\end{equation}
\begin{equation}
A_3 = \frac{\tv_a\tv_c e}{\tD^2}be\,,\ \gamma_3 =
      \frac{(2b)^2 \pDe}{\tv_a^2}\,. \nonumber
\end{equation}
It follows that the magnetoresistance along the \vc\ is given by:
\begin{eqnarray} \label{eq:MRzzBab}
\rho_{zz} & \approx & \exp\left[ \beta\tD\sqrt{1+A_3|B|(\sncn{3})^{1/2}} \right]  \nonumber \\
& &    \times \left( 1 + C_3 A_3|B|\ssncn{3} \right)\,.
\end{eqnarray}

\begin{figure}
\includegraphics*[width=7cm]{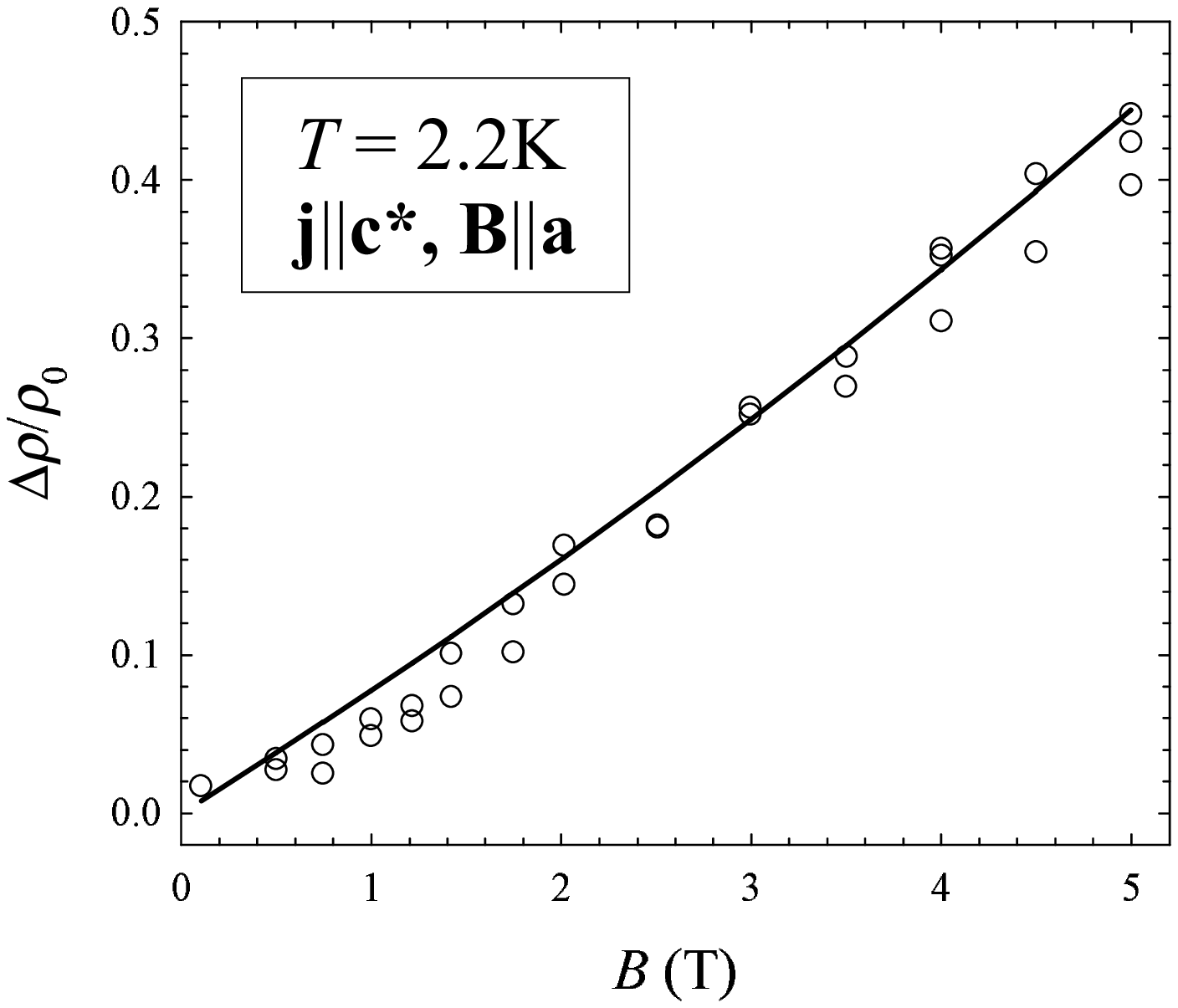}
\caption{\label{fig:MRzzBab} Magnetic field dependence of $\MR$ at 2.2~K for \jc\ and \Ba. Solid line is
fit to the theory (see text).}
\end{figure}
\begin{figure}
\includegraphics*[width=7cm]{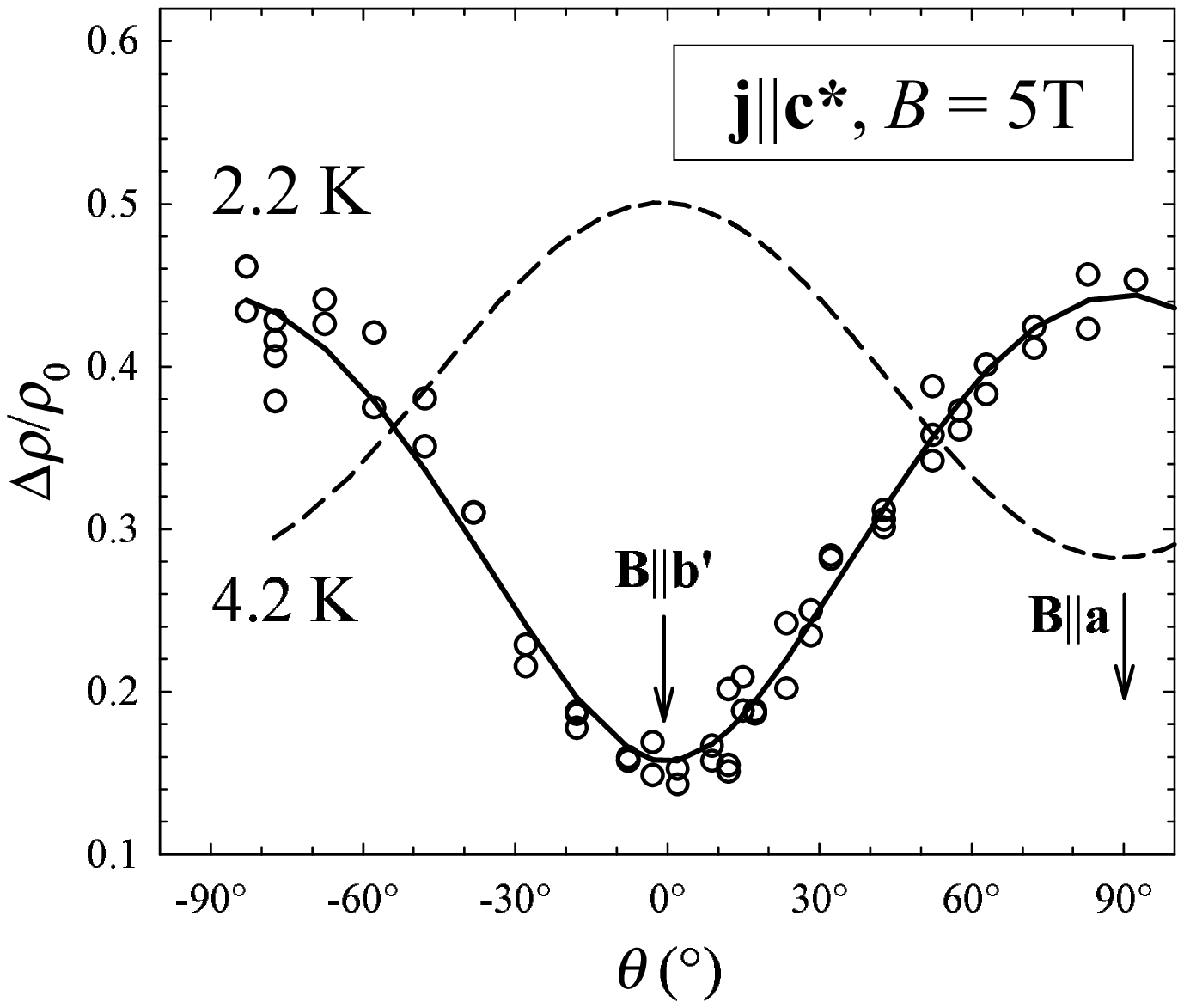}
\caption{\label{fig:MRzzThab} Angular dependence of $\MR$ at 2.2~K and 4.2~K, $B=5$~T, for \jc, \vB\ in
\abplane\ plane. Solid line is fit to the theory (see text).}
\end{figure}
Figure \ref{fig:MRzzBab} presents the magnetic field dependence of MR for \jc\ and \Ba\ at 2.2~K, while
\myreffig{fig:MRzzThab} shows the angular dependence of magnetoresistance for \jc, $B=5$~T at 2.2~K (see
\myreffig{fig:config}, case \cs{C}). We point out that there is a maxima in MR for \Ba\ at 2.2~K, while
there is a minima in MR for \Ba\ at 4.2~K (dashed line \myreffig{fig:MRzzThab}). This kind of behaviour
cannot be described in terms of conventional SDW where the imperfect nesting plays the crucial role.
Namely, in that case we expect maxima in MR for \Bb. This big change in MR anisotropy may be described
within our new theoretical model. We shall compare now our experimental data at 2.2~K with the
\myrefeq{eq:MRzzBab}. The solid line is fit based on the theory that describes the data on
\myreffig{fig:MRzzBab} and \myreffig{fig:MRzzThab} at 2.2~K very well. We deduce $A_3=0.0165$~\Tm, $C_3
\approx 0$, and $\gamma_3=0.154$. From $A_3$ we obtain:
\[
    \frac{\tv_a\tv_c}{\tD^2} = \xi_a\xi_c = 1.087 \times 10^{-13}\text{~cm}^2
\]
and assuming  $\xi_c/\xi_a = 1/13.6$ we obtain  $\xi_a = 1.2\times 10^{-6}$~cm which is quite
reasonable.\cite{IshiguroBook98} On the other hand $\gamma_3=0.154$ gives $\Delta_1/\Delta=1.75$ that is too
large, at least by a factor of 2, giving $\Delta_1\approx 60$~K.

\section{Concluding remarks}
We have proposed that the phase transition at \Tfour\ in \PF\ is due to the appearance of the USDW in
addition to the already existing SDW. As we have shown, the quasiparticle spectrum in SDW with imperfect
nesting and/or USDW in a magnetic field is, due to the Landau quantization, very different from the one
for $B=0$. The appearance of USDW order parameter modifies the quasiparticle spectrum. This change is
readily accessible to both the magnetoresistance and the angular dependence of the magnetoresistance.
Indeed, USDW describes the dramatic change in the magnetoresistance below \Tfour. Furthermore, from the
angular dependence of the magnetoresistance we can deduce the parameters $\tD = 20$~K, $v_a/\Delta =
\xi_a =1.2\times 10^{-6}$~cm, and $v_c/v_a=7.33\times 10^{-2}$, which are consistent with the previously
known values. However, the new order parameter $\Delta_1$, associated with USDW, appears to behave
somewhat unexpectedly (as the deduced values give $\Delta_1=20$~K, 1~K, and 60~K for \vB\ in the
\acplane\ plane, in the \bcplane\ plane and in the \abplane\ plane, respectively). The reason for
differences of $\Delta_1$ is unclear at present. We note, however, that in contrast to our earlier
analysis,\cite{KorinHamzicEPL98} here we have taken into account the Landau quantization of the
quasiparticle spectrum, but we have considered the \vk -independence of the scattering rate. We can only
suppose, that in addition to the Landau quantization the inclusion of the \vk -dependent $\Gamma$ would
solve this  $\Delta_1$ discrepancy.

\begin{acknowledgments}
This experimental work was performed on samples prepared by K. Bechgaard. We thank P. Thalmeier for
drawing our attention to Ref.\ \onlinecite{NersesyanJLTP89} and \onlinecite{NersesyanJPCM91}. We
acknowledge useful discussions with A. Hamzi\'{c} and S. Tomi\'{c}.
\end{acknowledgments}

\bibliography{mbasletic}

\end{document}